\documentclass[letterpaper,10pt]{article} 

\usepackage{osameet3} 
\usepackage{braket}
\usepackage[square,sort,comma,numbers]{natbib}
\usepackage{amsmath,amssymb}

\usepackage[colorlinks=true,bookmarks=false,citecolor=blue,urlcolor=blue]{hyperref} 

\usepackage{subcaption}
\usepackage[export]{adjustbox}
\usepackage[font=footnotesize]{caption}

\begin{document}

\title{\vspace{-2cm}Inverse Design of Quantum Holograms in Three-Dimensional Nonlinear Photonic Crystals}

\vspace{-1cm}
\author{Eyal Rozenberg\textsuperscript{1}, Aviv Karnieli\textsuperscript{3}, Ofir Yesharim\textsuperscript{4}, Sivan Trajtenberg-Mills\textsuperscript{3}, Daniel Freedman\textsuperscript{2}, Alex M. Bronstein\textsuperscript{1} and Ady Arie\textsuperscript{4}}
\address{\textsuperscript{1}Department of Computer Science, Technion, Haifa, Israel \hspace{0.3cm} \textsuperscript{2}Google Research, Haifa, Israel\\
\textsuperscript{3}Raymond and Beverly Sackler School of Physics and Astronomy, Tel Aviv University, Israel\\
\textsuperscript{4}School of Electrical Engineering, Fleischman Faculty of Engineering, Tel Aviv University, Israel}
\email{eyalr@campus.technion.ac.il}

\maketitle
\vspace{-0.5cm}
\begin{abstract}
We introduce a systematic approach for designing 3D nonlinear photonic crystals and pump beams for generating desired quantum correlations between structured photon-pairs. Our model is fully differentiable, allowing accurate and efficient learning and discovery of novel designs.
\vspace{0.3cm}
\end{abstract}

\begin{figure*}[ht]
\captionsetup[subfigure]{aboveskip=0.5pt,belowskip=2pt}
    \centering
    \begin{subfigure}{0.47\linewidth}
        \centering
        \adjincludegraphics[width=\textwidth]{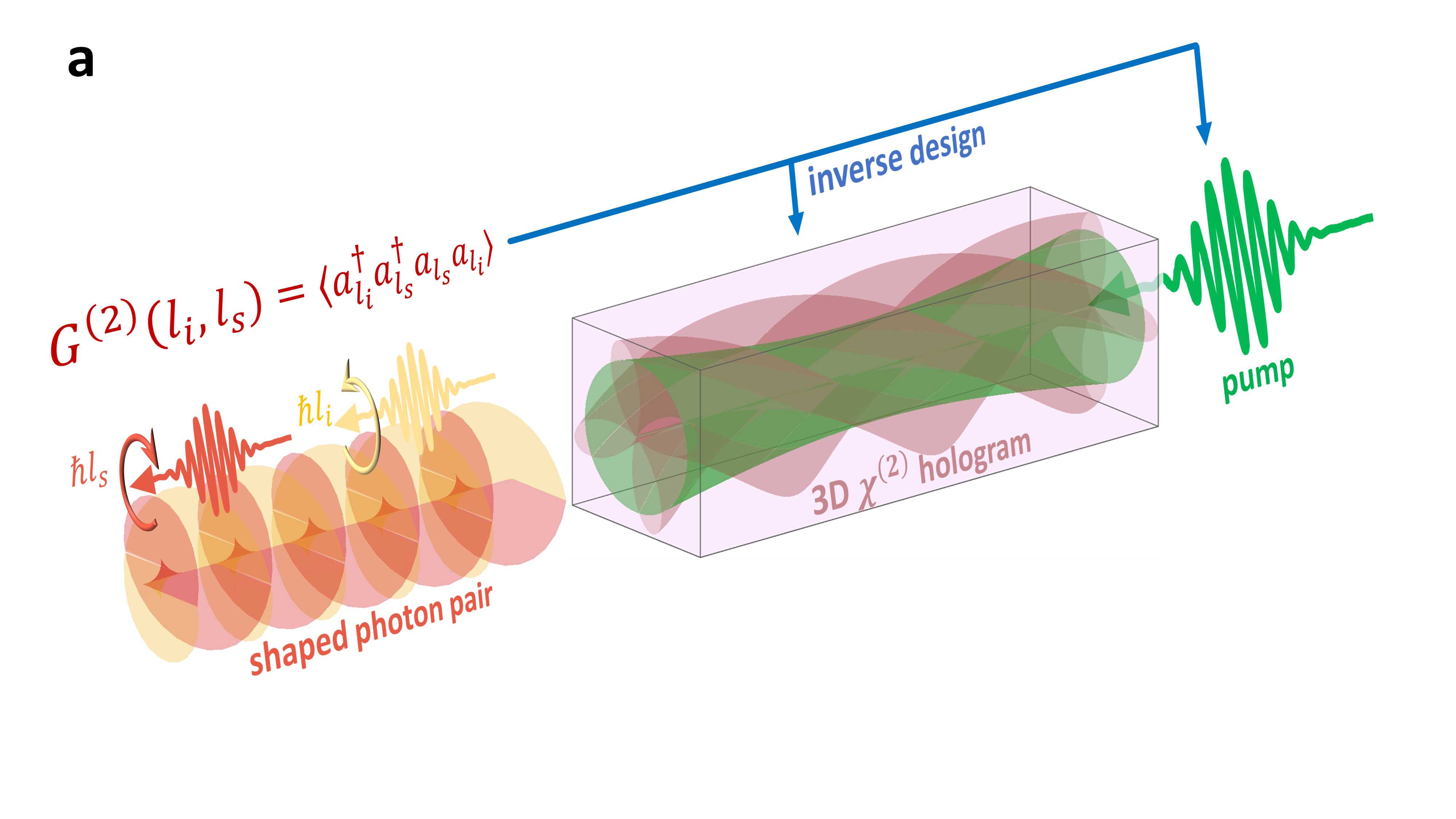}
    \end{subfigure}%
    \hspace{0.15em}%
    \begin{subfigure}{0.47\linewidth}
        \centering
        \adjincludegraphics[width=\textwidth]{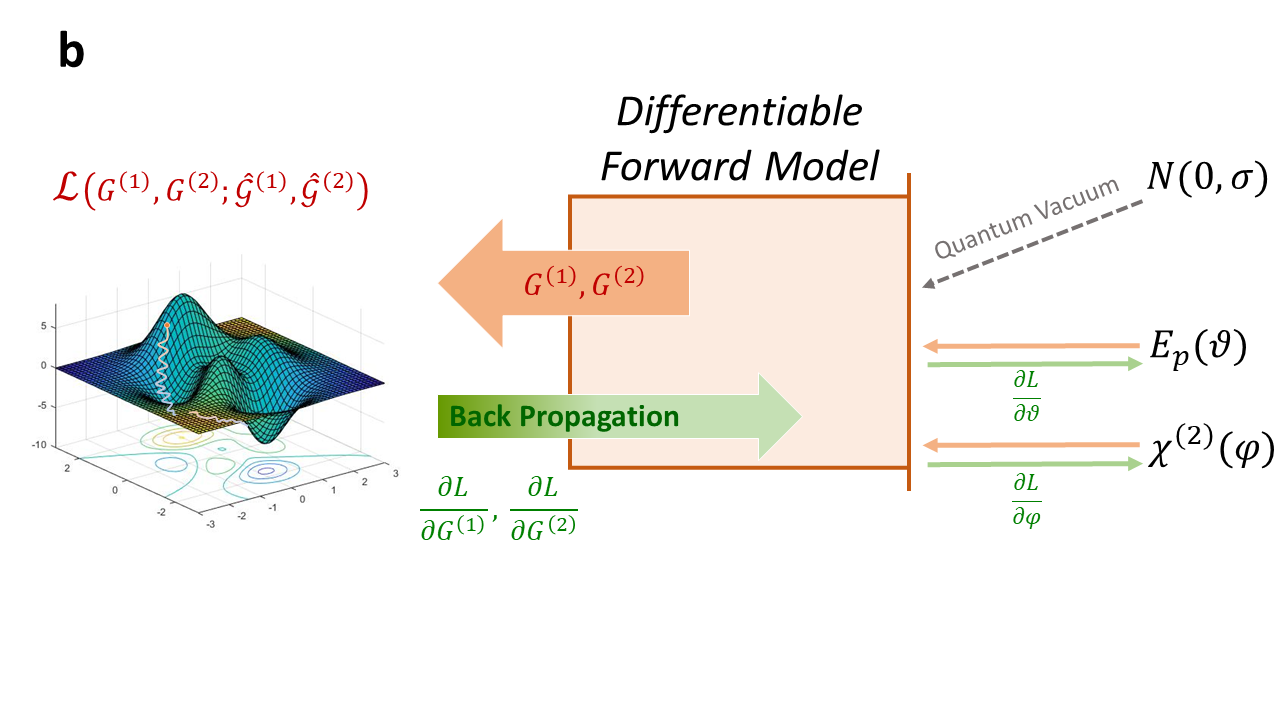}
    \end{subfigure}%
    \vspace{-0.75cm}
    \caption{\emph{Inverse design of quantum holograms model}. \hspace{0.1cm} \textbf{a)} A nonlinear photonic crystal hosts a three-dimensional hologram, and the pump field can be constructed from different spatial modes. The resulting down-converted shaped photon pairs display second-order quantum correlations. The inverse problem is defined by learning the crystal hologram and pump modes that result in the desired quantum correlations. \hspace{0.1cm} \textbf{b)} An illustration of the architecture of the model. The pump and crystal-hologram, \(E_p, \chi^{(2)}\), are parameterized by \(\vartheta, \varphi\) which are learned parameters of the model. The stochastic nature of the SPDC interaction is set by the \emph{quantum vacuum}, defined by a non-differentiable random node. The forward model simulates the light-matter interaction and results in the 1st and 2nd quantum correlations, \(G^{(1)}, G^{(2)}\). Their discrepancy against a desired correlation is measured via an appropriate loss function. The model is fully differentiable and is optimized using \(backpropagation\) and first order optimization method \emph{Adam}.}
\label{fig:concept}
\end{figure*}

Quantum optics has proven to be an invaluable resource for the realization of many quantum technologies, such as quantum communication, computing and cryptography. A prominent reason for this is the availability of sources generating nonclassical light, mainly based on nonlinear interactions; the most prevalent of which is spontaneous parametric down-conversion (SPDC) in second order nonlinear $\chi^{(2)}$ materials \cite{SPDCreview2018}. Many of these schemes focus on creating high-dimensional quantum states by employing the spatial degrees of freedom of free-space modes, such as Laguerre-Gauss (LG) beams carrying orbital angular momentum (OAM) \cite{Forbes2019review,kovlakov2018quantum,liu2018coherent} or Hermite-Gauss (HG) beams \cite{Walborn2012,KovlakovPRL2017}. Shaping of quantum correlations of entangled photons has received much attention, owing to possible application in quantum communication. However, their manipulation via the spatial mode of the pump beam is quite limited \cite{liu2018coherent,kovlakov2018quantum}. Recent breakthroughs in the fabrication of three-dimensional nonlinear photonic crystals \cite{XuNatPhot2018} offer a promising new avenue for shaping and controlling arbitrary quantum correlations through the design of three-dimensional $\chi^{(2)}$ holograms (see Fig.\ref{fig:concept}.a).

\vspace{-0.3cm}

\vspace{-0.5cm}
The task of finding the right hologram (modulation pattern of the crystal nonlinearity) to provide the desired quantum correlations falls into a broad family of \emph{inverse problems} in optics. 
Given a forward model which maps a hologram to the quantum correlations it produces, a loss function may be defined which measures the discrepancy between the quantum correlations produced by the forward model and the desired correlations.  A common approach to solving the inverse problem then involves minimizing the loss function, see Figure \ref{fig:concept}.
Although SPDC is a quantum mechanical process, simulation tools (i.e.,~forward models) nevertheless exist and have been shown to agree with experimental results \cite{brambilla2004simultaneous, lantz2004spatial, trajtenberg2020simulating}. Unfortunately, SPDC simulations are stochastic in nature, which makes it challenging to integrate them directly into optimization-based approaches to the inverse problem.  More specifically, such approaches require the ability to compute the derivative of the parameters being optimized with respect to the loss function, which is rendered difficult due to the stochasticity. Recent attempts to overcome this issue \cite{kovlakov2018quantum} were not gradient-based, and were only used to learn the pump modes, disregarding the crystal structure. 

In this work, we discover novel designs of three-dimensional quantum holograms in nonlinear photonic crystals for generating desired quantum correlations between structured photon pairs, for example those of high-order maximally-entangled OAM states - sought after for applications in quantum information. We obtain these designs by solving the quantum optical inverse problem (Fig.\ref{fig:concept}) using efficient inference and learning in the directed SPDC probabilistic model, by employing the \emph{reparameterization} approach \cite{kingma2013auto}. Our implementation allows the gradients of the loss function to be directly calculated with respect to the physical parameters of the model, and to thereby be used in the optimization process. In contrast to previous realizations, we do not need to differentiate with respect to non-deterministic parameters. 

\begin{figure*}[ht]
\centering
\begin{tabular}{cc}
    \includegraphics[width=0.5\linewidth]{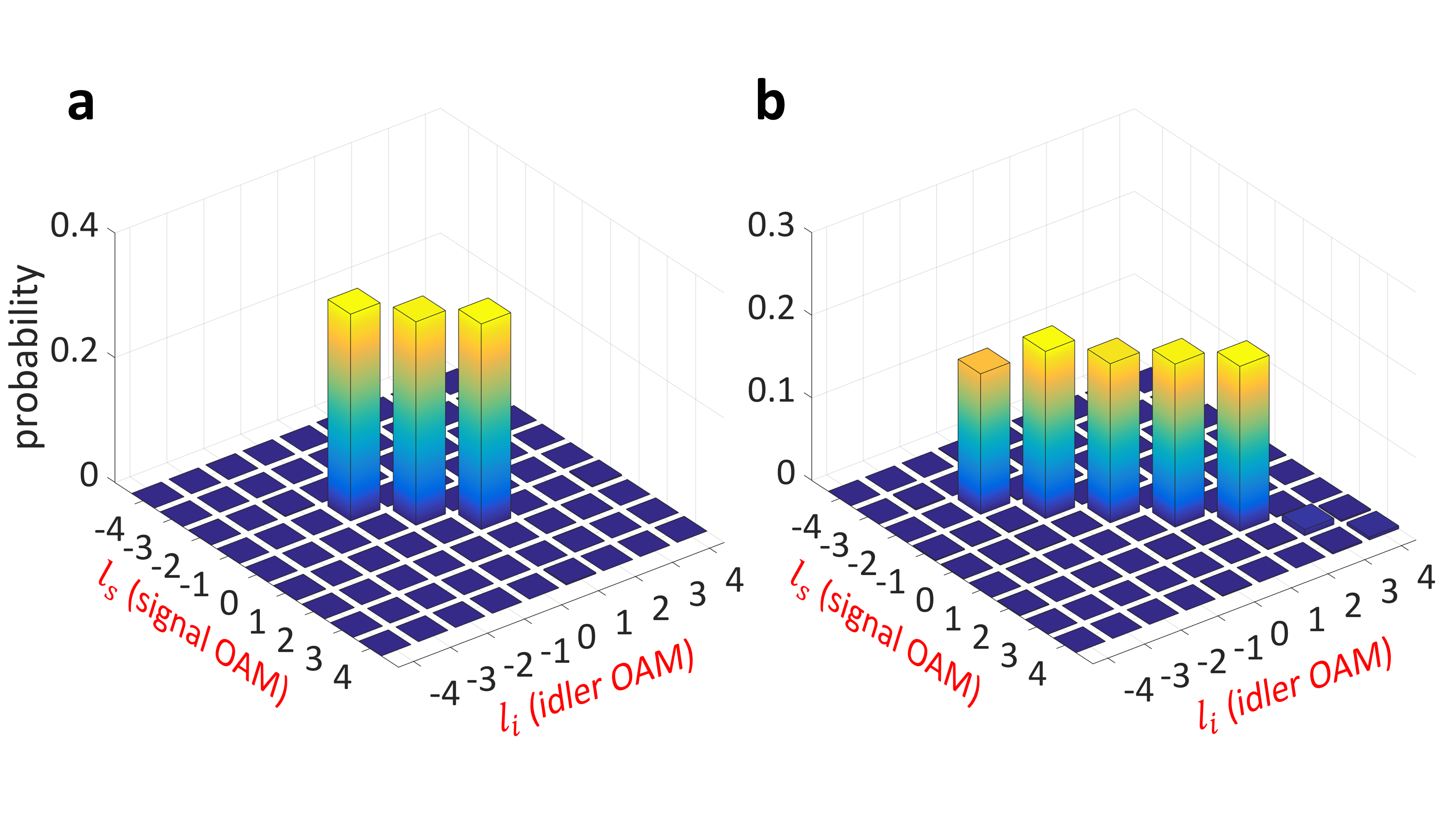}
    \includegraphics[width=0.5\linewidth]{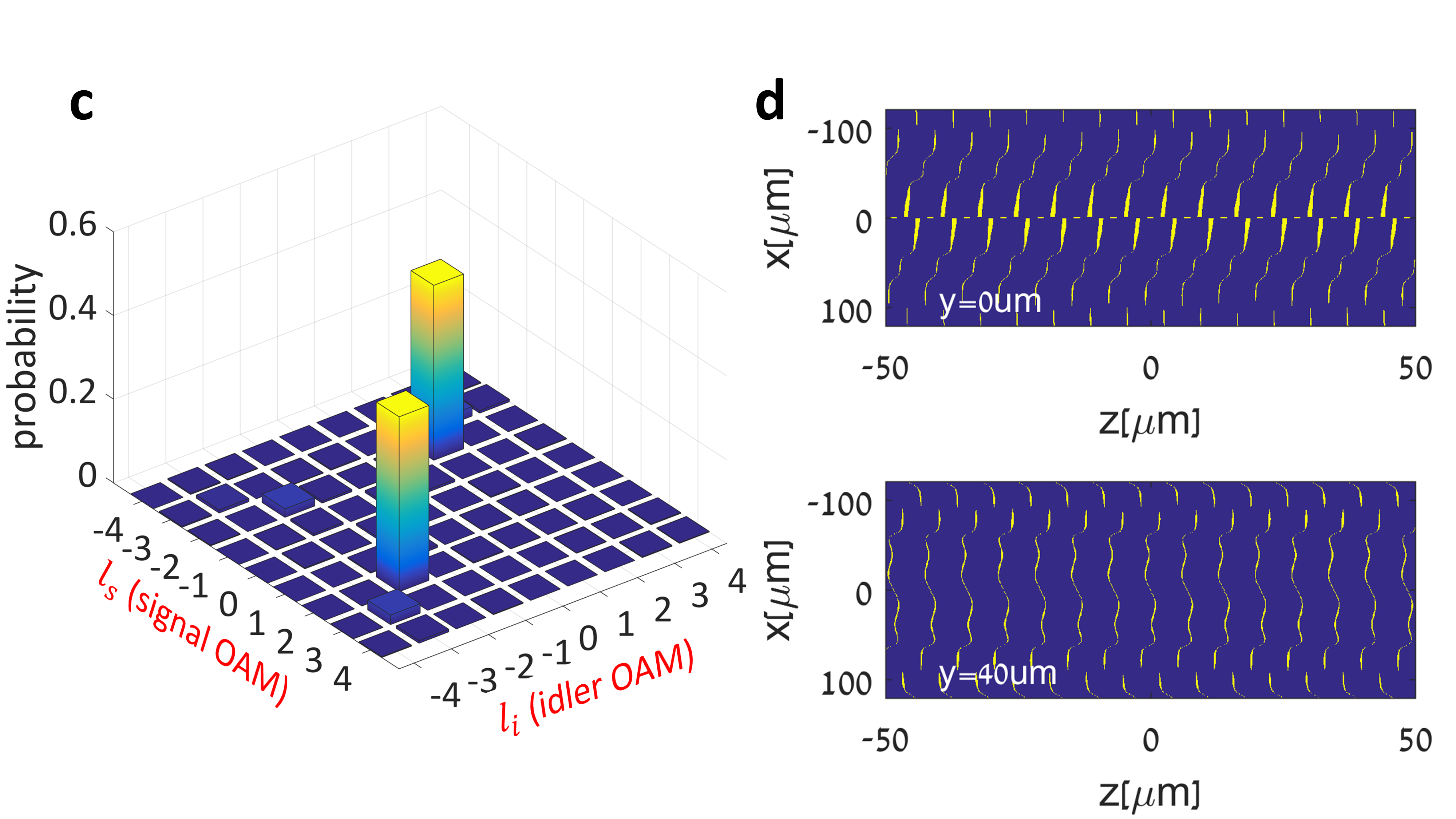}
\end{tabular}
\vspace{-0.5cm}
 \caption{Results of learning quantum correlations in the Laguerre-Gauss basis. \textbf{a-c} Learned two-photon quantum correlations of \hspace{0.1cm}\textbf{a)} qutrit ($d=3$) \hspace{0.1cm}\textbf{b)} ququint ($d=5$) and \hspace{0.1cm}\textbf{c)} high-order qubit ($l=2$). \hspace{0.1cm}\textbf{d)} Learned three-dimensional crystal pattern for the qubit case (panel \textbf{c}) for two cross-sections in the $y$ coordinate. (Crystal patterns producing \textbf{a-b} not shown due to space constraints.)}
 \label{fig:lg_results}
\end{figure*}

The results for learning desired quantum correlations of photon pairs in the LG basis are visualized in Fig.\ref{fig:lg_results}. As a target for learning, we provide the algorithm with a desired quantum state of the two-photon probability distribution derived from the second-order quantum correlation $P(l_i,l_s) = G^{(2)}(l_i,l_s,l_s,l_i)=\braket{\psi|a^{\dagger}_{l_i}a^{\dagger}_{l_s}a_{l_s}a_{l_i}|\psi}$; where $\ket{\psi}$ denotes the quantum state, $a$ ($a^{\dagger}$) denote the photon annihilation (creation) operators, and $l_i,l_s$ denote the OAM quantum number of the idler and signal photons. We perform simulations for a 1mm long $\mathrm{LiNbO_3}$ crystal, and with a 532nm, 1mW CW pump. The process is type-II quasi-phase-matched for an on-axis generation of polarization-distinguishable photon pairs at 1064nm. First, we let the algorithm learn a three-dimensional hologram inducing the quantum correlations associated with OAM qudit states, with $d=3$ (Fig.\ref{fig:lg_results}a) and $d=5$ (Fig.\ref{fig:lg_results}b), of the form $\ket{\psi}=\sum_{l=1}^d \mathrm{e}^{i\varphi_l} \ket{l,l}/\sqrt{d}$. In addition, we also learn correlations of high-order OAM qubits of the form $\ket{\psi}=(\ket{l,-l}+\mathrm{e}^{i\varphi}\ket{-l,l})/\sqrt{2}$, with $l=2$ (Fig.\ref{fig:lg_results}c-d), where we also show the learned 3D crystal patterns.

\vspace{-0.4cm}
\begin{figure*}[ht]
\centering
\begin{tabular}{cc}
    \includegraphics[width=0.5\linewidth]{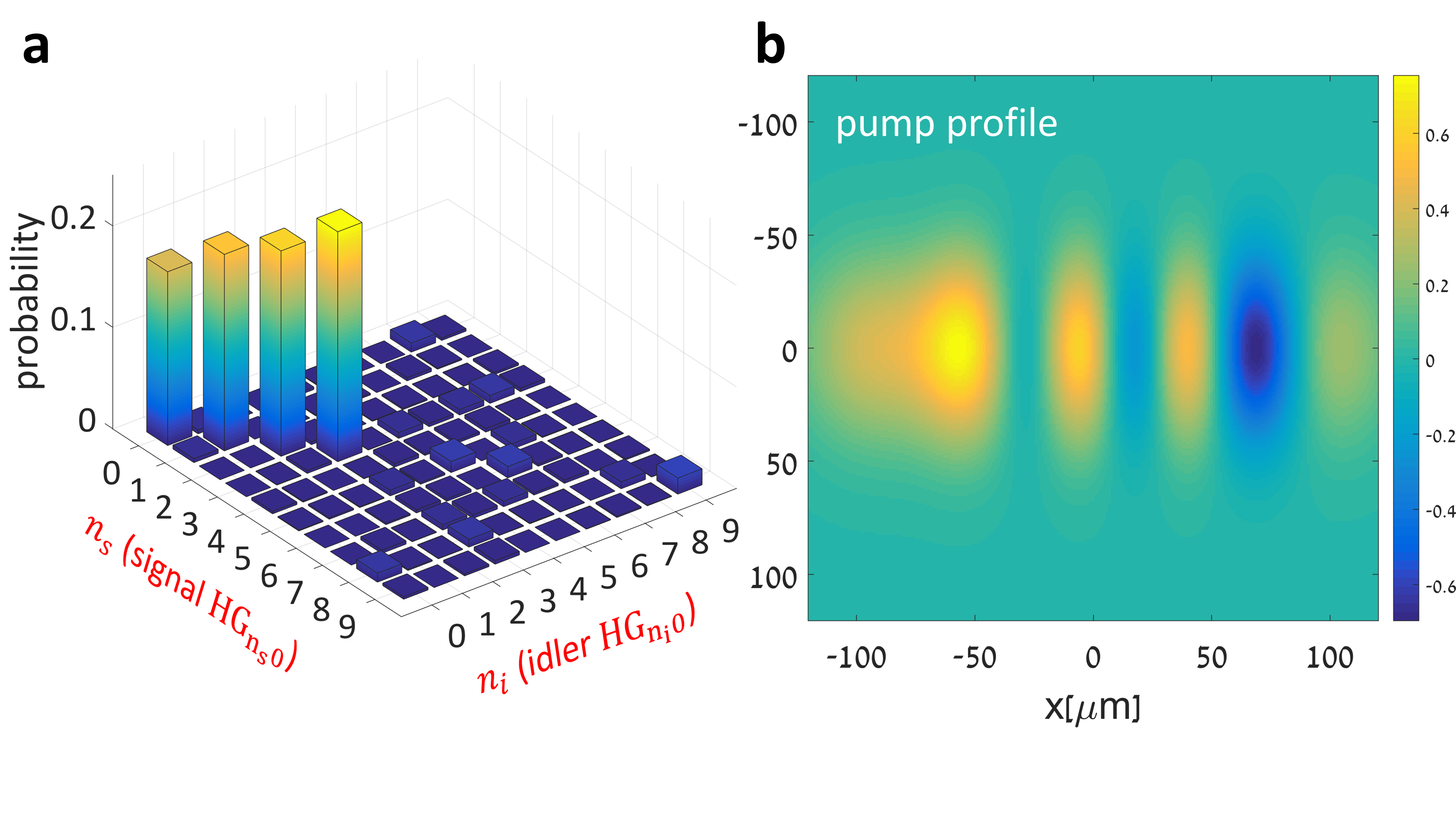}
    \includegraphics[width=0.5\linewidth]{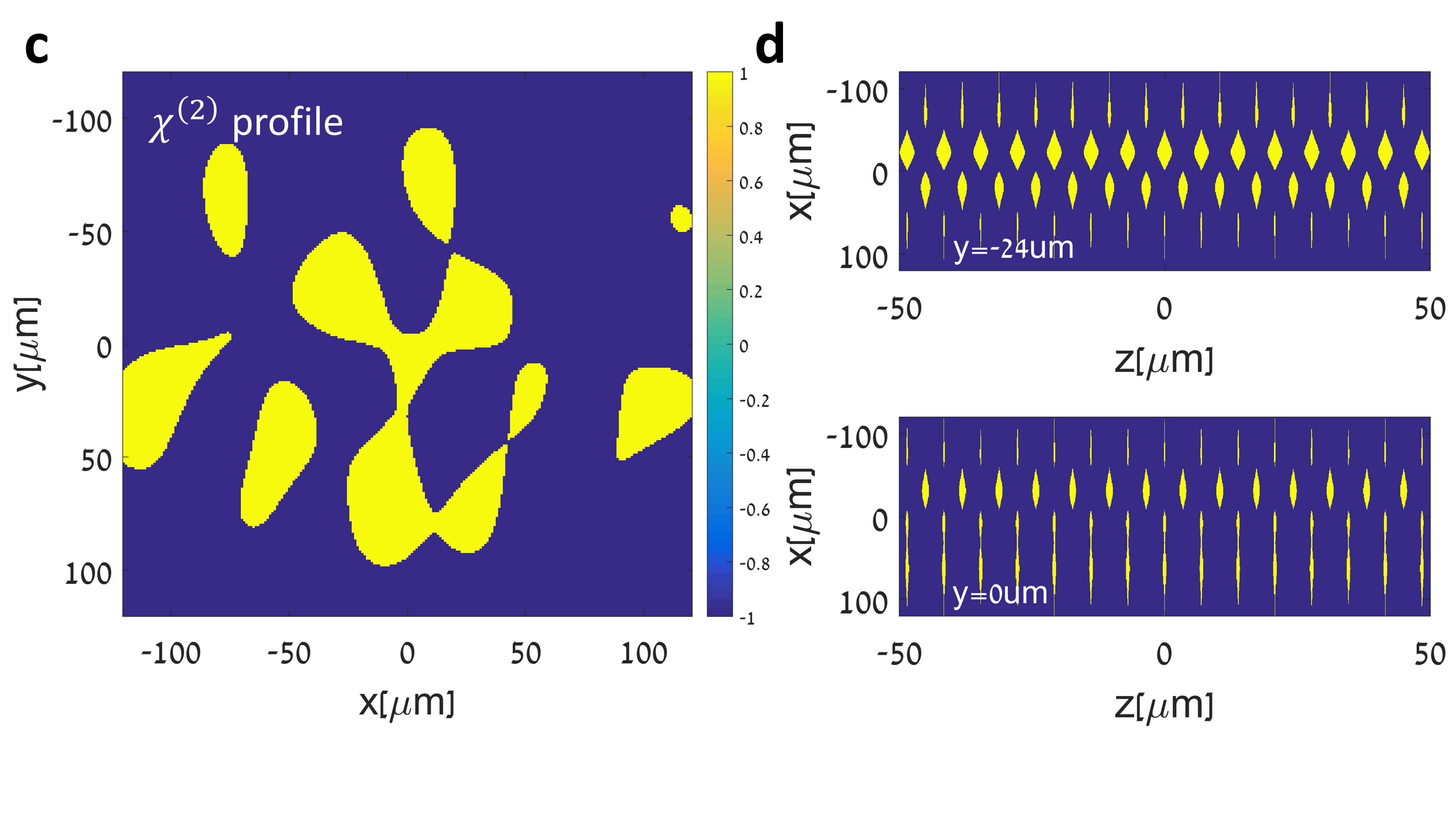}
\end{tabular}
\vspace{-0.7cm}
 \caption{Results of learning quantum correlations in the Hermite-Gauss basis. \textbf{a)} Learned correlations of a ququad ($n_s, n_i$ are the quantum mode numbers of the signal and idler). \textbf{b)} Learned pump magnitude (a.u.). \textbf{c)} Learned crystal profile at $z=0$. \textbf{d)} Learned crystal pattern along $z$.}
 \label{fig:hg_results}
\end{figure*}

\vspace{-0.5cm}
Similarly to the LG case, we now turn to the simultaneous learning of both the pump \textit{and} crystal patterns, which will allow us to achieve desired correlations in the HG basis. We let the algorithm learn the quantum correlations of a maximally-entangled HG ququad ($d=4$) of the form $\ket{\psi}=1/2(\ket{\mathrm{HG}_{00}}\ket{\mathrm{HG}_{00}}+\ket{\mathrm{HG}_{10}}\ket{\mathrm{HG}_{10}}+\ket{\mathrm{HG}_{20}}\ket{\mathrm{HG}_{20}}+\ket{\mathrm{HG}_{30}}\ket{\mathrm{HG}_{30}})$. The results of this learning of pump and crystal structure are depicted in Fig. \ref{fig:hg_results}.

\vspace{-0.35cm}
\vspace{-0.6cm}

To conclude, we have shown the inverse design of three-dimensional nonlinear photonic crystals and pump beam profiles for shaping quantum correlations between spatial modes of SPDC photon pairs, for several high-dimensional quantum states used in quantum information. The run time was about 15 minutes on an NVIDIA TITAN Xp 12Gb GPU. We intend to extend our work to allow for learning of the full density matrix of a target quantum state, as well as the longitudinal variation of the nonlinear crystal controlling the joint spectral amplitude.

\vspace{-0.2cm}
\bibliography{references}

\begin{thebibliography}{10}
\newcommand{\enquote}[1]{``#1''}

\bibitem{SPDCreview2018}
C.~Couteau, \enquote{Spontaneous parametric down-conversion,}
  {\protect\JournalTitle{Contemporary Physics}} \textbf{59}, 291 (2018).

\bibitem{Forbes2019review}
A.~Forbes and I.~Nape, \enquote{Quantum mechanics with patterns of light:
  Progress in high dimensional and multidimensional entanglement with
  structured light,} {\protect\JournalTitle{AVS Quantum Sci.}} \textbf{1},
  011701 (2019).

\bibitem{kovlakov2018quantum}
E.~V. Kovlakov \emph{et~al.}, \enquote{Quantum state engineering with twisted
  photons via adaptive shaping of the pump beam,}
  {\protect\JournalTitle{Physical Review A}} \textbf{98}, 060301 (2018).

\bibitem{liu2018coherent}
S.~Liu \emph{et~al.}, \enquote{Coherent manipulation of a three-dimensional
  maximally entangled state,} {\protect\JournalTitle{Physical Review A}}
  \textbf{98}, 062316 (2018).

\bibitem{Walborn2012}
S.~P. Walborn and A.~H. Pimentel, \enquote{Generalized hermite–gauss
  decomposition of the two-photon state produced by spontaneous parametric down
  conversion,} {\protect\JournalTitle{J. Phys. B}} \textbf{45}, 165502 (2012).

\bibitem{KovlakovPRL2017}
E.~V. Kovlakov \emph{et~al.}, \enquote{Spatial bell-state generation without
  transverse mode subspace postselection,} {\protect\JournalTitle{Phys. Rev.
  Lett.}} \textbf{118}, 030503 (2017).

\bibitem{XuNatPhot2018}
T.~Xu \emph{et~al.}, \enquote{Three-dimensional nonlinear photonic crystal in
  ferroelectric barium calcium titanate,} {\protect\JournalTitle{Nat. Phot.}}
  \textbf{12}, 591 (2018).

\bibitem{brambilla2004simultaneous}
E.~Brambilla \emph{et~al.}, \enquote{Simultaneous near-field and far-field
  spatial quantum correlations in the high-gain regime of parametric
  down-conversion,} {\protect\JournalTitle{Physical Review A}} \textbf{69},
  023802 (2004).

\bibitem{lantz2004spatial}
E.~Lantz \emph{et~al.}, \enquote{Spatial distribution of quantum fluctuations
  in spontaneous down-conversion in realistic situations,}
  {\protect\JournalTitle{The European Physical Journal D-Atomic, Molecular,
  Optical and Plasma Physics}} \textbf{29}, 437--444 (2004).

\bibitem{trajtenberg2020simulating}
S.~Trajtenberg-Mills \emph{et~al.}, \enquote{Simulating correlations of
  structured spontaneously down-converted photon pairs,}
  {\protect\JournalTitle{Laser \& Photonics Reviews}} \textbf{14}, 1900321
  (2020).

\bibitem{kingma2013auto}
D.~P. Kingma and M.~Welling, \enquote{Auto-encoding variational bayes,}
  {\protect\JournalTitle{arXiv preprint arXiv:1312.6114}}  (2013).

\end{thebibliography}
\bibliographystyle{osajnl.bst}

\end{document}